\documentclass[12pt,preprint]{aastex}

\usepackage{graphicx}
\usepackage{epstopdf}
\DeclareGraphicsExtensions{.eps}

\slugcomment{Submitted to Physics Education May 19, 2015}

\begin{document}

\title{Demonstrating Martian Gravity} 

\author{Patrik Pirkola and Patrick B. Hall}
\affil{Department of Physics and Astronomy, York University, Toronto, Canada}

\begin{abstract}

The surface gravity on Mars is smaller than the surface gravity on Earth, resulting in longer falling times.
This effect can be simulated on Earth by taking advantage of air resistance and buoyancy, which cause low density objects to 
fall slowly enough to approximate objects falling on the surface of Mars. 
We describe a computer simulation based on an experiment that approximates Martian gravity, and verify our numerical results by performing the experiment.
%
\end{abstract}


\section{Introduction}

The gravitational acceleration $g$ at the surface of an approximately
spherical planet is $g=GM/R^2$, where $G$ is the gravitational constant
and $M$ and $R$ are, respectively, the mass and radius of the planet.


The time $T$ required to fall from a height $h$ due to only
a gravitational acceleration $g$ 
can be derived from the equation for distance travelled in time $T$
at a fixed acceleration:
\begin{equation}
h = \frac{1}{2}gT^2 ~~\longrightarrow~~ T(h) = 
\sqrt{2h/g}.
\end{equation}
When the acceleration varies (due to air resistance),
a computer simulation can be used to calculate the fall time $T$.

We have designed an experiment on Earth to match the fall time of an object on Mars.  The MarsSim row in Table 1 gives the fall time results of a computer simulation using the parameters of that experiment.  A drop height of 2 meters matches the fall time on Mars and produces a relatively good match to Martian dynamics. 

\section{Logistics}
\subsection{Theory}
The net acceleration at time $t$
on a falling object of total mass $m_{\rm tot}$
is a downward (negative) gravitational acceleration 
plus upward air resistance and buoyancy terms:
\begin{equation}
a(t) = -g + \frac{1}{2}\rho_{\rm air} C_{\rm drag} A [v(t)]^2 /m_{\rm tot} + F_{\rm buoy}/m_{\rm tot}
\end{equation}
where the density of air is $\rho_{\rm air}$,
the drag coefficient of the object is $C_{\rm drag}$,
the horizontal cross-sectional area of the object is $A$,
and the object's downward velocity is $v(t)$.
The buoyancy force is $F_{\rm buoy}=m_{\rm air}g$, 
where $m_{\rm air}$ is the mass of air displaced by the object.
\subsection{The Simulation}
For our simulation, the object is a hollow rectangular box or bag of 
vertical depth $d$, length $\ell$ and width $w$ (so that $A=\ell w$),
with sides of negligible thickness and with mass $m_{\rm obj}$.
For such an object, 
the mass of air inside the box is 
$m_{\rm air}=\rho_{\rm air}\ell w d$ and
the total mass of the falling box is $m_{\rm tot}=m_{\rm obj}+m_{\rm air}$.
%
The drag coefficient for a rectangular box will lie between 
$C_{\rm drag}=1.05$ (cube, $\ell=w=d$) and $C_{\rm drag}=1.17$ 
(thin rectangular plate with $\ell/w<5$ and $d\ll\ell$; Hoerner 1965).
We assume $C_{\rm drag}=1.11$ for our rectangular box with 
$\ell\simeq w$ and $d/\ell\simeq 0.5$.
Using the above quantities in the expression for $a(t)$, the velocity $v(t)$ 
and height fallen $z(t)$ can be found numerically starting from
$ a(0)=-g $, $ v(0)=0 $ and $ z(0)=h $.
The time $T$ required to fall a distance $h$ is found by
stopping the calculation when $z(T)=0$.

We use an empty, rectangular, vinyl bag
(Richards Homewares Clear Vinyl Jumbo Blanket Bag No.~441W)
with $\ell=0.533$\,m, $w=0.635$\,m and $d=0.279$\,m,
so that with $\rho_{\rm air}=1.225$ kg\,m$^{-3}$, $m_{\rm air}=0.116$\,kg.
The bag itself weighs $m_{\rm obj}=0.140$\,kg.
%
When spray-painted to resemble a Martian rock 
using a can of orange or reddish paint 
the bag weighs $m_{\rm obj}=0.180$\,kg.
\\With $\rho_{\rm air}=1.225$ kg\,m$^{-3}$, $C_{\rm drag}=1.11$, $A$ =  0.3385 m$^{2}$, $m_{\rm tot}$ =  0.2960 kg, and $F_{\rm buoy}$ = 1.1348 N,
we can then write (2) as
\begin{equation}
a(t) = -g + b[v(t)]^2  + a_{\rm buoy}
\end{equation}
where $b$ = 0.7774 m$^{-1}$ and $a_{\rm buoy}$ = 3.834 m\,s$^{-2}$.
\section{Results}
Table 1 gives the calculated times for the spray-painted bag to fall 
distances of 1, 2 and 3 meters on Earth.  
Those times match the times required for an object to fall
from the same heights on Mars to within $\pm$11\%.
The fall time from 2 meters was experimentally verified to be
$1.06 \pm 0.17$ seconds using 
3 observers' measurements of 3 separate drops. Since this drop was the best match to Martian gravity, we include its computer simulated dynamics as compared to real Martian gravity as Figure 1. We also include an image of the bag as Figure 2.
\begin{table}[!h]
  \caption{Gravitational Accelerations and Fall Times}
\smallskip
  \begin{tabular}{@{}lcccc@{}}
\hline
Planet & ~$g$ (m/s$^2$)~ & ~$T$~(1\,m)~ & ~$T$~(2\,m)~ & ~$T$~(3\,m)~ \\
\hline
Earth & 9.81 &  0.452 & 0.639 & 0.782 \\
Mars*  & 3.70 &  0.735 & 1.040 & 1.274 \\ 
MarsSim & ... & 0.656 & 1.038 & 1.404 \\ 
Moon  & 1.63 &  1.108 & 1.567 & 1.919 \\	
\hline
\end{tabular}\\
\label{tab:1}

*Also Mercury, which has nearly identical g.
\end{table}
\begin{figure}[!h]
 \includegraphics[scale=0.4]{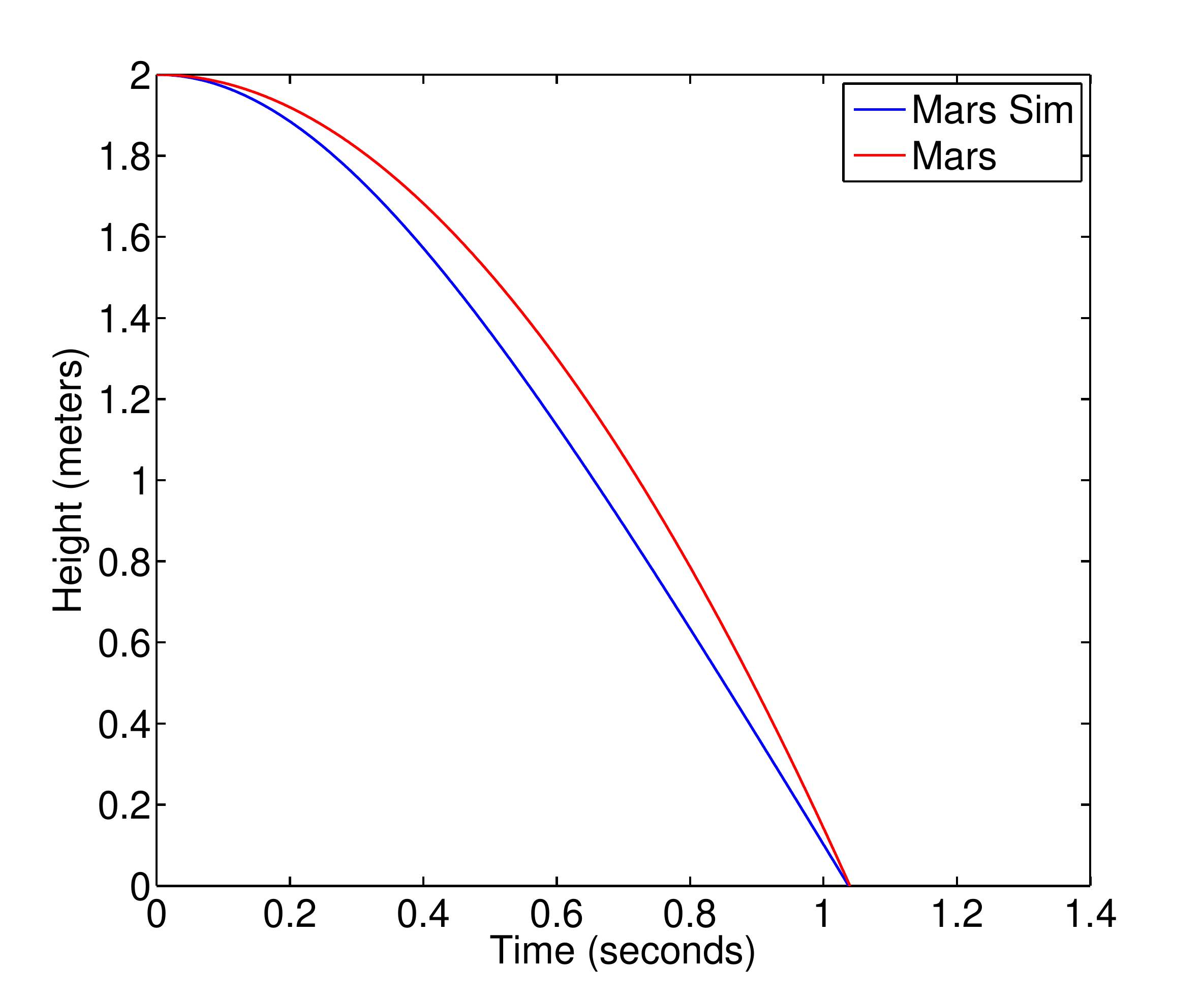}

 \caption{Two meter Mars simulation versus Martian drop.}
\end{figure}
\begin{figure}[!h]
 \includegraphics[scale=0.2]{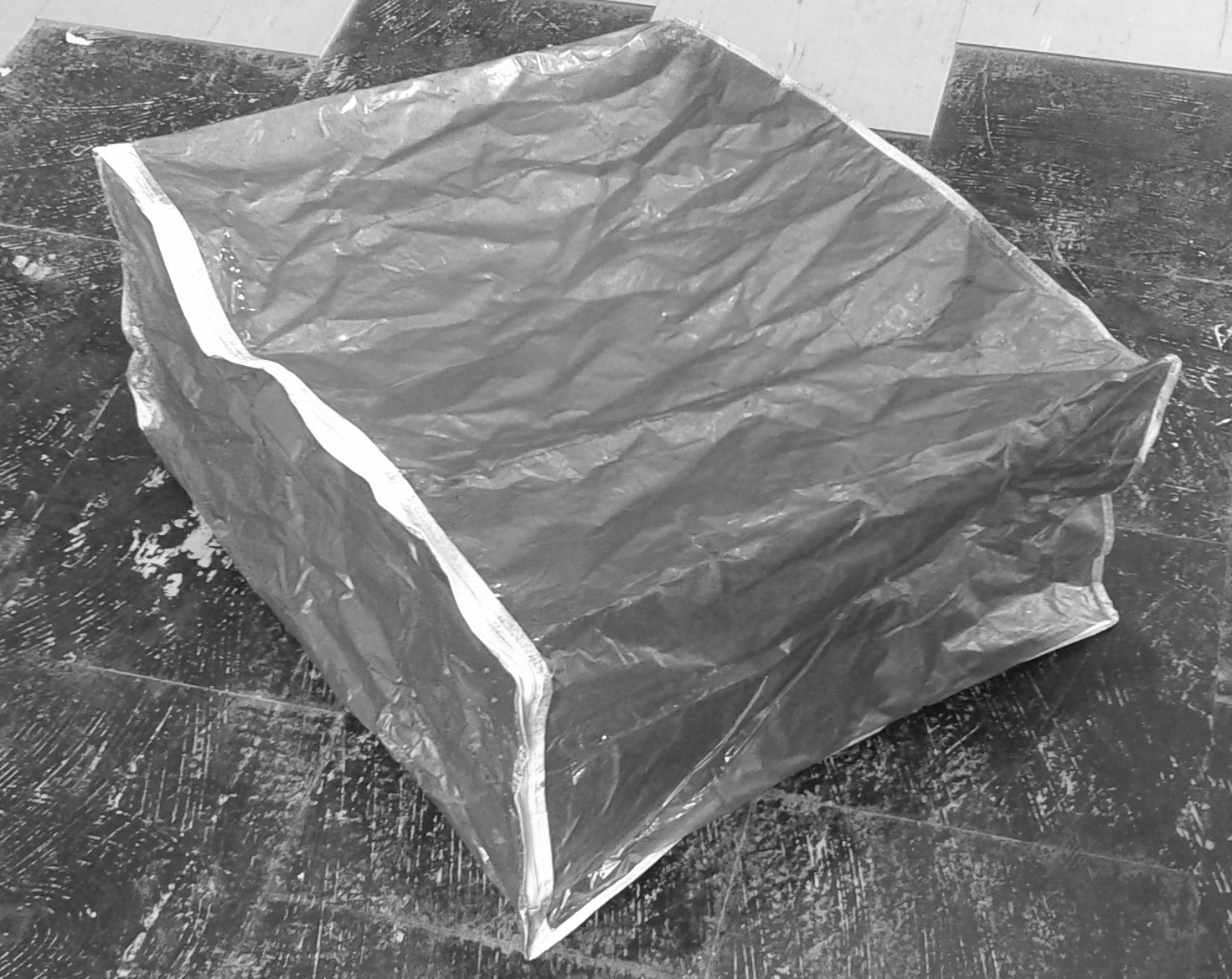}

 \caption{The bag used for the experiment.}
\end{figure}
\section{Conclusion}
This simulation can be used as a straightforward demonstration 
of the dynamics in Martian gravity as compared
to Earth gravity, by simultaneously dropping from the same height, an identical bag filled with textiles.

This simulation can also be used to illustrate the concept of 
air resistance.  Rotating the bag can yield different values of $A$,
resulting in different fall times.  
Furthermore, fall time measurement distributions, averages and
root-mean-square uncertainties can be calculated 
by having students time one or more drops
and record their measurements.

Alternatively, this simulation can be presented as a computational
assignment.  For example, given an object of fixed volume and mass (matching 
the spray-painted bag), find the horizontal surface area it must have so that the fall time is the same on Earth as on Mars for a given height.

~

~





\section*{References}
\begin{enumerate} 
\item Hoerner, S. F. 1965, Fluid Dynamic Drag, 2nd edition (Bakersfield, CA: Hoerner Fluid Dynamics), 3-16
\end{enumerate}


\end{document}